\documentclass[conference]{IEEEtran}
\usepackage{cite}
   \usepackage[pdftex]{graphicx}
\usepackage[cmex10]{amsmath}
\usepackage{amssymb}
\begin{document}
\title{Schema Redescription in Cellular Automata: Revisiting Emergence in Complex Systems}

\author{\IEEEauthorblockN{Manuel Marques-Pita}
\IEEEauthorblockA{School of Informatics and Computing\\ 
Indiana University (USA) and\\
Instituto Gulbenkian de Ci\^{e}ncia (Portugal)\\
Email: marquesm@indiana.edu
}
\and
\IEEEauthorblockN{Luis M. Rocha}
\IEEEauthorblockA{School of Informatics and Computing\\
Indiana University (USA) and\\
Instituto Gulbenkian de Ci\^{e}ncia (Portugal)\\
Email: rocha@indiana.edu}}


%Telephone: (800) 555--1212, Fax: (888) 555--1212}


\maketitle

\begin{abstract}
%\boldmath

We present a method to eliminate redundancy in the transition tables
of Boolean automata: \emph{schema redescription} with two symbols.
One symbol is used to capture redundancy of individual input
variables, and another to capture permutability in sets of input
variables: fully characterizing the \emph{canalization} present in
Boolean functions.
Two-symbol schemata explain aspects of the behaviour of
automata networks that the characterization of their emergent patterns
does not capture.
We use our method to compare two well-known cellular automata rules for the
\emph{density classification task} \cite{Mitchell:1996fk}: GKL
\cite{Gacs:1978kx,Gonzaga-de-Sa:1992vn} and GP \cite{Andre96b}. 
We show that despite having very different emergent behaviour, these
rules are very similar. Indeed, GKL is a special case of GP.
Therefore, we demonstrate that it is more feasible to compare cellular
automata via schema redescriptions of their rules, than by looking
at their emergent behaviour, leading us to question the tendency in
complexity research to pay much more attention to emergent patterns
than to local (micro-level) interactions.

\end{abstract}


% \fi
\IEEEpeerreviewmaketitle

\section{Introduction}

The intersection of biology and computer science has been fertile for
some time. Indeed, von Neumann was a member of the mid-twentieth
century Cybernetics group \cite{heims_cybernetics_group}, whose main
focus was the understanding of natural and artificial systems in terms
of communication and control. Most early computer science developments
were inspired by the models of cognition that orbited this group
\cite{McCulloch:1943uq}.
Since then, the need to understand how biological systems are able
to control and transmit information throughout the huge number of
components that comprise them has only increased. 
The study of complex network dynamics has received much attention in
the last two decades.  From pioneering work on networks of automata
\cite{Kauffmann:1969fk,Derrida86,Kauffmann:1993fk} to recent models of
genetic regulation
\cite{Mendoza:1998kx,Albert:2003ij,Espinosa-Soto:2004kc,Kauffman:2003ys},
it is clear that to understand and control the biological
organization, it is useful to study models based on complex networks
of automata \cite{Chaves2005,Bornholdt:2005xi,Bornholdt:2008kl}.

There has been much progress in understanding the structure of natural
networks---at the level of their topology
\cite{Barabasi:2002uq,Newman:2006fk} or of their more fine-grained
motifs \cite{Alon2007a}---as well as on modelling biological systems
as networks of automata.
Yet, we are still to fully grasp \emph{how} complex networks
``compute'' and how to harness them to perform specific tasks
\cite{Mitchell:2006fk}.
The need for a better understanding of collective computation in
complex networks has been identified in many areas.
For instance,  the way plants adjust stomatal
apertures for efficient gas exchanges on leaf surfaces is
statistically indistinguishable from the dynamics of automata that
compute \cite{Peak:2004fk}, and the high degree of connectivity in
biochemical intracellular signal transduction networks endows them
with the capability of emergent nontrivial classification---via
collective computation \cite{Helikar:2008fk}.
However, not much work in automata models of complex
networks has focused on how interactions at the local level of
components are linked to collective behaviour.

% It is often the case that, when studying a complex system, researchers
% build simulations of, or gather data on, the interactions between its
% constituent elements.
% %
% Afterwards, statistical inference and graph theoretical methods are
% often used to identify groupings, or temporal cascades, of such
% interactions in the networks thus inferred. 
% %
% These can reveal key interactions that at the right time control the
% collective behaviour of the complex system.
% %

We show that eliminating redundancy in the transition tables of local
automata reveals the loci of control of collective dynamics, better
than via the characterization of collective behaviour, emergent
patterns, and the like.
Our mechanism to eliminate redundancy is \emph{schema redescription}
\cite{Marques-Pita:2008cr, Marques-Pita:2008kl}, which we expand here
with an additional symbol as explained below.
Schemata remove redundancy from automata transition tables, leaving
only input states, or groups of input states that \emph{control}
state-transitions, known as their \emph{canalizing inputs}
\cite{Reichhardt:2007bs}.
Moreover, we discuss how schema redescription can reveal the
links between local and collective behaviour in automata networks
(here cellular automata).
We exemplify our approach with the classical problem of collective
computation in cellular automata: the \emph{density classification
  task} \cite{Mitchell:1996fk}.

\section{Cellular Automata}
\label{sec:emerg-comp-cell}

A $\kappa$\emph{-state automaton} is an integer variable $x \in
\{0,...,\kappa-1\}$, whose value (or state) is computed at time $t+1$
via a state-determined \emph{transition function} $F$, which takes as
input the state of a set of $n$ automata at time $t$.
When $x=0$ we refer to it as being in the \emph{quiescent} state; any
other state is regarded as an \emph{active} state.
To study the dynamics of systems of many simple elements (known as
complex systems), it is common to study networks of interconnected
automata.
One way to do this is to arrange automata in a regular lattice of $N$
cells, which we refer to as \emph{cellular automata}, a
formalization of a complex dynamical system with its origins in the
work of Ulam and von Neumann \cite{vonNeumann66}.
In a homogeneous cellular automaton (CA), which we use here, each cell
is defined by the same $\kappa$-state automaton and is connected to
$n$ neighbour cells in the lattice (including itself). Therefore,
there are $\kappa^n$ possible \emph{local neighbourhood
  configurations} (LNCs).
The transition function for every cell, $F=\{f_j\}$, can thus be
defined as a \emph{look-up table} (LUT) where each entry, $f_j, j \in
\{1, ..., \kappa^n\}$, is an assignment of a possible input LNC,
or\emph{ condition part}, to an output state $x_i(t+1) \in
\{0,...,\kappa-1\}$.
In binary CAs ($\kappa=2$) it is possible to classify individual cell state
transitions in three categories:
(1) \emph{preservations,} where a cell does not change its state in
the next time step, i.e.  $x_i(t) = x_i(t+1)$;
(2) \emph{generations,}  in which the cell goes from the
quiescent to the active state;
and (3) \emph{annihilations,} where the cell goes from the active to
the quiescent state.
The \emph{initial configuration} (IC) of states of a CA lattice is
typically random.
In the CAs we consider here, cells in the lattice update their states
synchronously.
%
%The execution of a CA for a number $M$ of discrete time steps from a
%given IC, where cells typically update their states synchronously and
%there are periodic boundary conditions, is represented as the set
%containing $M+1$ lattice state configurations.

%LMR: We never use the symbol M in the paper, so the phrase above is not needed.

\section{Schema Redescription}
\label{sec:canal-bool-autom}

We previously proposed \emph{schema redescription} with one symbol as
a method to relate an automaton's LUT to the collective dynamics it
produces in one- and two-dimensional CA lattices
\cite{Marques-Pita:2008cr,Marques-Pita:2008kl}. Here we introduce
two-symbol schemata. 
The basic idea is simple: LUTs are compressed into parsimonious
\emph{schemata} with two extra symbols that capture redundant input
states.
One symbol captures the \emph{irrelevance} of some inputs in some situations;
the other the \emph{permutability} of subsets of inputs.
Only the input states that effectively determine the automaton's state
transitions remain.
A related notion of \emph{canalizing inputs} exists in the context of
modelling biochemical networks, e.g. those that regulate the
expression patterns of genes. Indeed, automata with canalizing inputs
have been shown to stabilize the dynamics of Boolean models of genetic
regulation
\cite{Kauffman:2003ys,Grefenstette:2006rv,Reichhardt:2007bs}.

\subsection{Wildcard Schemata}
\label{sec:schemata}

\begin{figure}[!ht]
\centering
\includegraphics[width=3.6in]{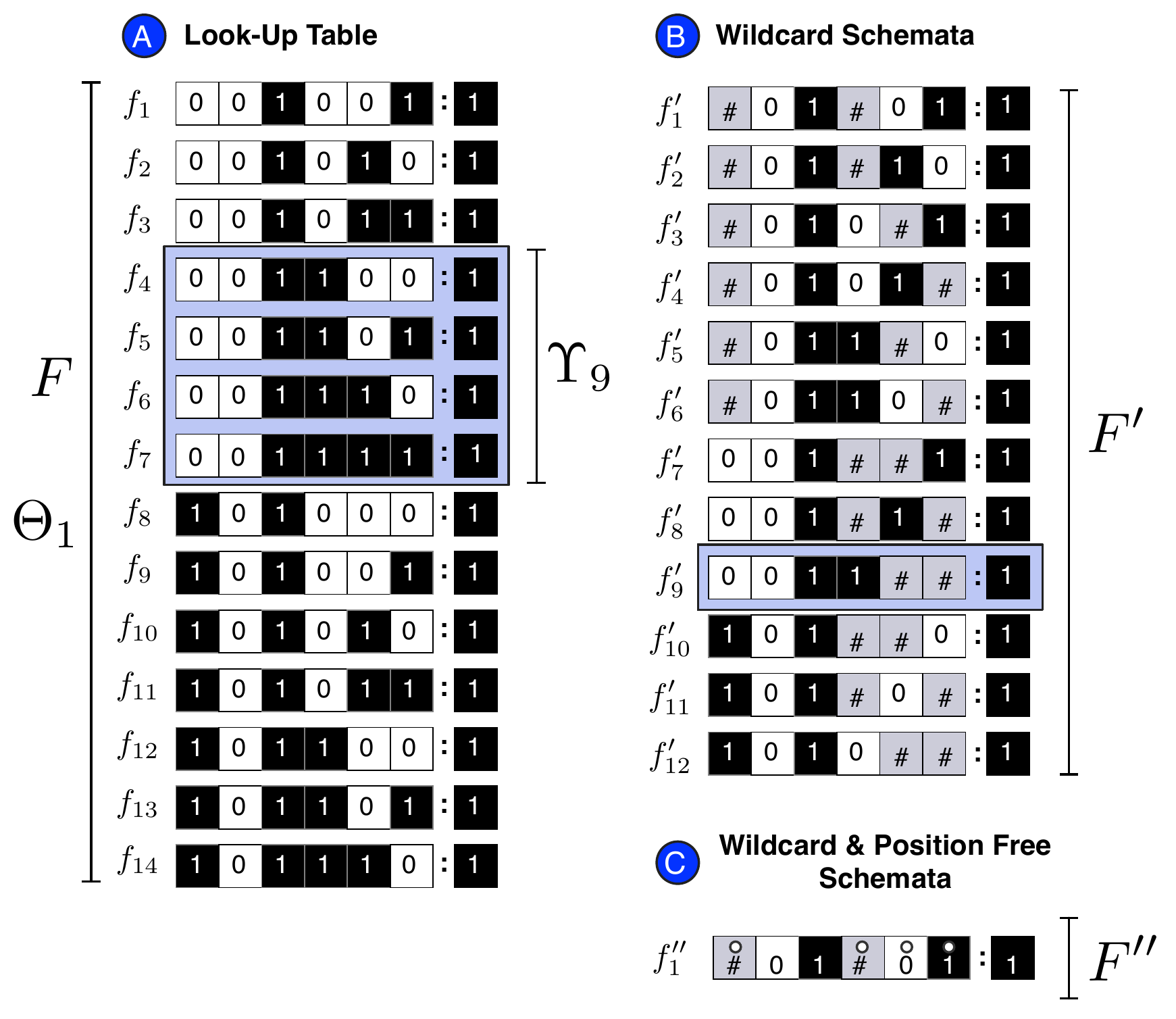}
\vspace*{-.15in}
\caption{A set of LUT entries with transition to 1, redescribed. White
  (black) states are 0 (1). Wildcards are grey.  Notice that
  $\Upsilon_9 \equiv \{f_4,f_5,f_6,f_7\}$.  Using the additional
  position-free symbol, the entire set $F'$ is compressed into a
  single two-symbol schema: $f''_1$. Any permutation of the inputs
  with the position-free symbol results in a schema in $F'$.  Since
  there is only one set of marked inputs, the position-free symbol
  does not require an index. }
\label{sch_example}
\vspace*{-.1in}
\end{figure}

To unpack the notion of schema, consider a Boolean automaton $x$,
whose state-transition function depends on the states of six Boolean
inputs (possibly other automata).
Figure \ref{sch_example}A depicts the subset of LUT entries
($f_{\alpha}$) for automaton $x$, that lead to state-transition
$x(t+1) =1$.

We first compress LUT entries using schemata that are like LUT
entries, but allow an additional \emph{wildcard} $(\#)$ symbol
(represented graphically in grey) to appear in their condition part.
A wildcard input means that \emph{any valid state
  is accepted for it, with no effect on the state-transition}.
This results in a redescription of the set of LUT entries, $F$, for an
automaton $x$ of $n$ inputs into a smaller set of wildcard schemata,
$F' \equiv \{f'_{\upsilon}\}$ (see Figure \ref{sch_example}B). 
The set $F'$, for a given automaton also contains the original entries
in its LUT, $F$, that could not be redescribed by wildcard schema. 
Each specific wildcard schema $f'_{\upsilon}$ redescribes a subset of
entries in the original LUT, denoted by $\Upsilon_{\upsilon} \equiv
\{f_{\alpha}: f_{\alpha} \rightarrowtail f'_{\upsilon}\}$
($\rightarrowtail$ means ``is redescribed by'').
Wildcard schemata are \emph{minimal} in the sense that none of the
(non-wildcard) inputs in the condition part of a schema can be
``raised'' to the wildcard status and still ensure the automaton's
transition to the same state.
Because wildcard schemata are minimal, it follows that for any two
wildcard-schemata $f'_\upsilon$ and $f'_\phi$, $\Upsilon_{\upsilon}
\nsubseteq \Upsilon_{\phi} \wedge \Upsilon_{\phi} \nsubseteq
\Upsilon_{\upsilon}$.
Because wildcard schemata are \emph{minimal} and \emph{unique}, they
are equivalent to the set of \emph{prime implicants} obtained during
the first step of the Quine \& McCluskey Boolean minimization
algorithm \cite{Quine:1955ec}. 
In other words, a schema is \emph{unique} in the sense that the subset
of LUT entries it redescribes is not fully redescribed by any other
schema.
However, in general $\Upsilon_{\upsilon} \cap \Upsilon_{\phi} \neq
\emptyset$. This means that schemata can overlap in terms of the LUT
entries they describe.
In Figure \ref{sch_example}, $\Upsilon_1
\equiv \{f_1,f_5,f_9,f_{13}\}$ and $\Upsilon_9 \equiv
\{f_4,f_5,f_6,f_7\}$, therefore $\Upsilon_1 \cap \Upsilon_9 \equiv
\{f_5\}$.

Our redescription methodology derives from the work of John
Holland on condition/action rules to model inductive reasoning in
cognitive systems \cite{Holland:1986ly}.
However, the same idea had been developed previously for the
minimization of circuits in electrical engineering
\cite{Quine:1955ec}.
It was also used by Valiant \cite{Valiant:1984fk} when introducing
\emph{Probably Approximately Correct} (PAC) learning.

A wildcard schema can always be expressed as a logical conjunction of
literals (logical variables or their negation).
Since such a schema is a \emph{prime implicant}, it follows that all
of its literals are \emph{essential} to determine the automaton's
state transition.
We refer to the literals (input states) in a wildcard schema as its
\emph{essential input states} or \emph{enputs} for short.
For instance, schema $f'_9$ of the example of figure \ref{sch_example}
has 4 enputs: the first 4 cells in its condition part.
The set $F'$ can be expressed as a logical \emph{disjunctive normal
  form} (DNF)---that is, a disjunction of conjunctions.

\subsection{Two-Symbol Schemata}
\label{sec:2schemata}

Here we introduce a further redescription of $F'$ that results in a
set of two-symbol schemata $F'' \equiv \{f''_{\theta}\}$ (Figure
\ref{sch_example}C).
The \emph{position-free symbol} ($\circ_m$) above inputs in
a condition part means that \emph{any pair of inputs thus marked can
  ``switch places'' without affecting the automaton's
  state-transition.}
The index of the position-free symbol, when necessary, is used to
differentiate among distinct subsets of inputs whose elements can only
switch places with each other.
A two-symbol schema $f''_{\theta}$ redescribes a set $\Theta_{\theta} \equiv
\{f_{\alpha}: f_\alpha \rightarrowtail f''_{\theta}\}$ of LUT entries
of $x$. 
In the same way a wildcard schema captures a subset
$\Upsilon_\upsilon$ of LUT entries, a two-symbol schema redescribes a
subset of wildcard schemata \footnote{Given a two-symbol schema, $f''_{\theta}$, the wildcard schemata it
redescribes can be denoted by the set $\Theta'_\theta$. 
It is useful to look at set $\Theta'_\theta$ as a matrix where rows
correspond to wildcard schemata, and columns to their inputs.
The last column is the state transition of $f''_{\theta}$,
and may therefore be omitted.
The wildcard-schemata rows of matrix $\Theta'_\theta$ have the same
number of zeroes, ones and wildcards.
For any pair of inputs $(i,j)$ in $f''_{\theta}$ that are marked with
the same position-free symbol, the corresponding $i^{th}$ and $j^{th}$
columns of $\Theta'_\theta$ can be switched, and the rows of the
resulting matrix reordered to obtain $\Theta'_\theta$ again.}.
Therefore $f''_{\theta}$ captures the degree of \emph{symmetry} or
\emph{group invariance} in the set of wildcard schemata contained in
$\Theta'_\theta$ \cite{MCCLUSKEY:1956nx}.
We refer to the subsets of such marked inputs in a two-symbol schema as
\emph{group-invariant enputs}, which may include wildcard symbols
marked with the position-free symbol.
Since each two-symbol schema compresses a number of wildcard schemata,
it follows that for a given automaton $x$, $|F''| \leq |F'| \leq |F| =
2^k$.

After redescription, all redundancy is removed in the form of the two
symbols.
In a given schema, a wildcard input alone is not essential by
definition: it is never an \emph{enput}, while an individual input set
up to a Boolean state always is. 
However, when a wildcard input is marked with a position-free symbol,
then it is part of a group-invariant enput defined by the set of inputs
thus marked. 
In this case, the enput identifies a group invariance property, such
as \emph{``as long as two of these inputs are false''}.
To summarize, the set of wildcard schemata $F'$ of an automaton $x$
captures its enputs, while the addition of the position-free symbol
captures group invariant relationships in subsets of wildcard
schemata. 
In our working example, the resulting two-symbol schema (Figure
\ref{sch_example}) reveals that the automaton's transition to \emph{true} is
determined only by a subset of its six inputs:
\emph{as long as inputs 2 and 3 are \emph{false} and \emph{true},
  respectively, and among the others at least one is \emph{true} and
  another is \emph{false}, the automaton will transition to \emph{true}}.
Note that such minimal input constraints are not obvious by looking at
the LUT of the example automaton.

In previous work, we used wildcard schemata to identify \emph{conceptual
  properties} of known solutions to a well-studied problem that
requires non-trivial collective computation:
the \emph{density classification task} in homogeneous cellular
automata. By uncovering these properties, we were able to devise a
genetic algorithm that operated on a search space of wildcard
schemata, rather than the space of LUTs. 
We found that the space of schemata is much more amenable to search,
with very good correlations among good solutions to the problem. 
This way, we found some of the best solutions to this problem
\cite{Marques-Pita:2008kl,Marques-Pita:2008cr}.
Our aims here are: (1) introduce two-symbol schemata; (2) demonstrate
that they are more useful to explain and control the function and
behaviour of CAs than existing techniques to analyse collective
behaviour, and (3) it is more feasible to compare automata transition
functions via their schemata redescriptions, than by looking at their
emergent or collective behaviour.

\section{The Density Classification Task (DCT)}
\label{sec:dens-class-task}

The \emph{density classification task} (DCT) is the most studied
example of collective computation in CAs.
The goal is to find a binary CA that can classify the
majority state in a random IC.
If the majority of cells in the IC is in the quiescent (active)
state, after a number of time steps the lattice should converge
to a homogeneous state where every cell is in the quiescent (active)
state.
Since the outcome could be undecidable in lattices with even number of
cells $(N)$, lattices with an odd number of cells are used instead.
Devising CA rules that perform this task is not trivial, because cells in a
CA lattice update their states based on local neighbourhood
information, the states of 7 input Boolean variables in this case.
Therefore, this task requires information to be transferred across time
and (lattice) space to achieve a correct global classification.
The performance of CA rule $F$, on $C$ initial configurations of length $N$
is denoted by $\mathcal{P}^{C}_{N}(F)$.
The definition of the DCT used in our studies comes from
\cite{Mitchel93lambda}.

\section{Explanation in Complex Systems}
\label{sec:emergence}

\begin{figure}[t]
\begin{center}
\includegraphics[width=3.2in]{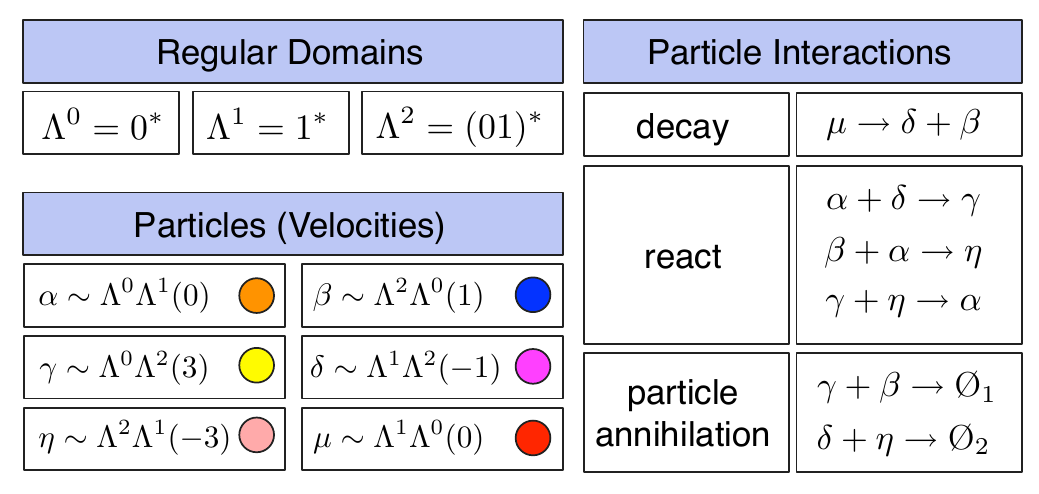}
\vspace{-.15in}
\caption{Catalogue of regular domains, particles (domain boundaries),
  particle velocities (in parentheses), and particle interactions seen
  in $F_{\textrm{GKL}}$ 's space-time behaviour. The colouring
  assigned to each particle is then used in Figure
  \ref{particle-ca-filtered}. The notation $p \sim \Lambda^x
  \Lambda^y$ means that $p$ is the particle forming the boundary
  between regular domains $\Lambda^x$ and $\Lambda^y$. (Adapted from
  \cite{Mitchell:1996fk}.)}
\label{particle-table}
\end{center}
\vspace{-.2in}
\end{figure}

The complex systems community has produced methodologies to
characterize space-time dynamics in CAs to study emergent behaviour,
often using the DCT as a test-case\cite{Mitchell:2006fk}.
The most well-known such methodology is the \emph{computational
  mechanics} framework (CM) \cite{Hanson:1993ux,Crutchfield:1994eh},
used to extract \emph{regular domains}: repeating patterns in the
space-time dynamics, formally defined as sets of regular languages.
It has been argued that in order to explain emergent behaviour, we
should look at the boundaries between domains, known as
\emph{particles}, as this is where information is exchanged in the CA
lattice \cite{Crutchfield:2003ui}.
Therefore, collective behaviour in CA dynamics is characterized by
finite catalogues of particles and their interaction rules, which
define the emergent computation implemented by a given CA
\cite{Mitchell:2006fk}, and can be seen to possess some features of
linguistic representation, though not in fully symbolic forms
\cite{rochahordijk05}.
Figure \ref{particle-table} depicts the catalogue of particles for the
$F_{\textrm{GKL}}$ rule\cite{Gacs:1978kx,Gonzaga-de-Sa:1992vn}, which
we describe below. 

Other methods exist to detect regularities in CA space-time dynamics,
typically statistical and information theoretical filters that uncover
patterns that correspond to the particles found via CM
\cite{Lizier:2008xs,Shalizi:2006lq}.
But when it comes to understanding and controlling collective
behaviour, a key issue is concerned with finding the best
\emph{explanatory} device to use.
Do we focus exclusively on the emergent patterns of collective
dynamics (the ``emergentist'' paradigm), as CM and related
methodologies do? Or can we gain explanatory advantages by paying more
attention to the lower level of local transition functions?

In the remaining of this article we pursue the second approach,
towards a \emph{hierarchical} methodology that seeks to link
local interactions to global behaviour, ultimately allowing us to
better understand and control the collective dynamics of complex
systems modelled as automata networks. \\

\section{Comparing Collective Dynamics}
\label{sec:comparing}

\begin{figure}[t]
\centerline{\includegraphics[width=4in]{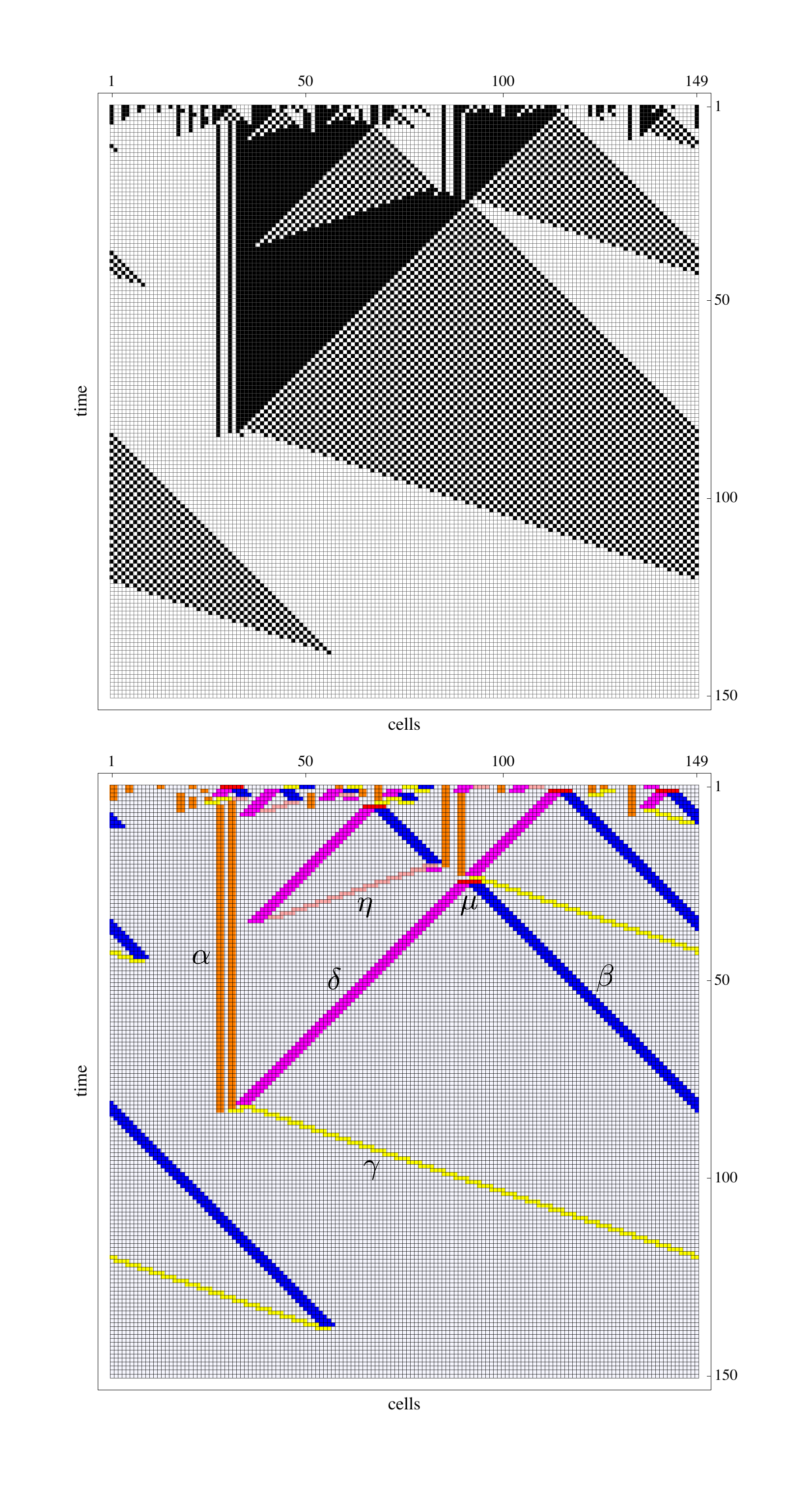}}
\vspace{-.35in}
\caption{\footnotesize{(Top) A space-time diagram produced by
    $F_{\textrm{GKL}}$ . (Bottom) The diagram with the regular
    domains filtered out, leaving only the locations of the  particles (coloured using
    the scheme in  Figure~\ref{particle-table}). Note that particle
$\alpha$ (red) lasts for only one time step,
after which it decays to particles $\delta$ and $\beta$.}}
\vspace{-0.2in}
\label{particle-ca-filtered}
\end{figure}

\begin{figure}[!ht]
\begin{center}
   \includegraphics[width=3.5in]{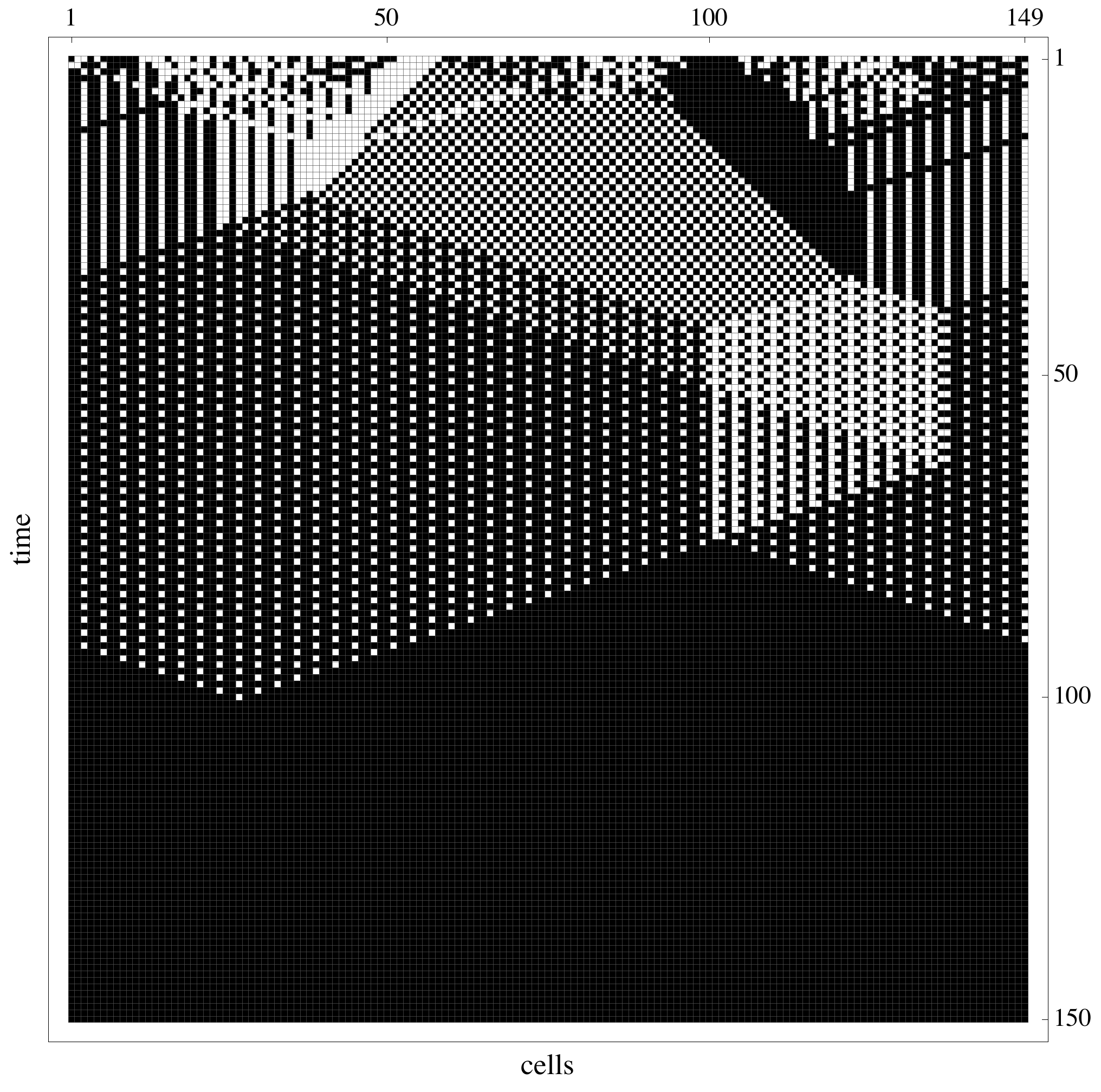}
   \caption{A space-time diagram produced by $F_{\textrm{GP}}$ . }
\label{koza-dyn}
\end{center}
\vspace{-.2in}
\end{figure}

Most of the known one-dimensional CA rules for the DCT were
analysed using wildcard schemata in \cite{Marques-Pita:2006vn}.
We found that most of these CAs are \emph{process symmetric}: they
classify equally well ICs with majority 0s and with majority 1s.
We also found that two of the most well-known high-performing CA rules
for the DCT, while observing very distinct collective behaviour, are
nonetheless similar at the local LUT level \cite{Marques-Pita:2006vn}.
More specifically, we refer to the human-derived CA $F_{\textrm{GKL}}$
\cite{Gacs:1978kx,Gonzaga-de-Sa:1992vn}, and $F_{\textrm{GP}}$,
derived via genetic programming \cite{Andre96b}.
Figures \ref{particle-ca-filtered} and \ref{koza-dyn} depict the
space-time dynamics of $F_{\textrm{GKL}}$ and $F_{\textrm{GP}}$,
respectively.
The performance of these rules on the DCT is:
$\mathcal{P}^{10^5}_{149}(F_{\textrm{GKL}}) \approx 0.815$, and
$F_{\textrm{GP}}$ $\mathcal{P}^{10^5}_{149}(F_{\textrm{GP}} ) \approx
0.822$.
In the remainder of this article, instead of $F_{\textrm{GKL}}$ we use
the fully equivalent $F_{\textrm{GKL}'}$, which is the mirror rule of
$F_{\textrm{GKL}}$.
That is, exactly the same CA rule, but where the particles traverse
the lattice in the opposite direction, with the same velocities and
interactions (see \emph{computationally equivalent} CAs in
\cite{Marques-Pita:2006vn}).

The collective behaviour of $F_{\textrm{GKL}'}$ and $F_{\textrm{GP}}$
is quite distinct; while $F_{\textrm{GKL}'}$, after a short transient,
leads to the appearance of three domains, the dynamics of the
$F_{\textrm{GP}}$ leads to the appearance of many more (we have
identified at least ten \cite{rochahordijk05}, see figure
\ref{koza-dyn}).
Because $F_{\textrm{GKL}'}$ only has three domains, from the CM
perspective, its collective dynamics can be characterized and
explained with a catalogue of six particles: the permutations without
repetition of two out of three possible domains, where order is
relevant\footnote{$p={d!}/{(d-2)!}$, where $d$ is the the number of
  domains.}.
Indeed, the collective behaviour of $F_{\textrm{GKL}}$ has been
well predicted by the CM framework, using only six particles
\cite{Mitchell:1994uq,HordijkEtAl1998}.
In contrast, $F_{\textrm{GP}}$ produces many more particles; even with
just ten domains (it has more), there would be ninety possible
particles and many potential interactions.
Interestingly, even though the collective dynamics of the two rules is
apparently (via the CM framework) completely different, we found
that $F_{\textrm{GP}}$ can be easily derived from $F_{\textrm{GKL}}$,
via a small sequence of operations on the wildcard schemata
representation of the latter.

\begin{figure}[!ht]
\begin{center}
   \includegraphics[width=2.9in]{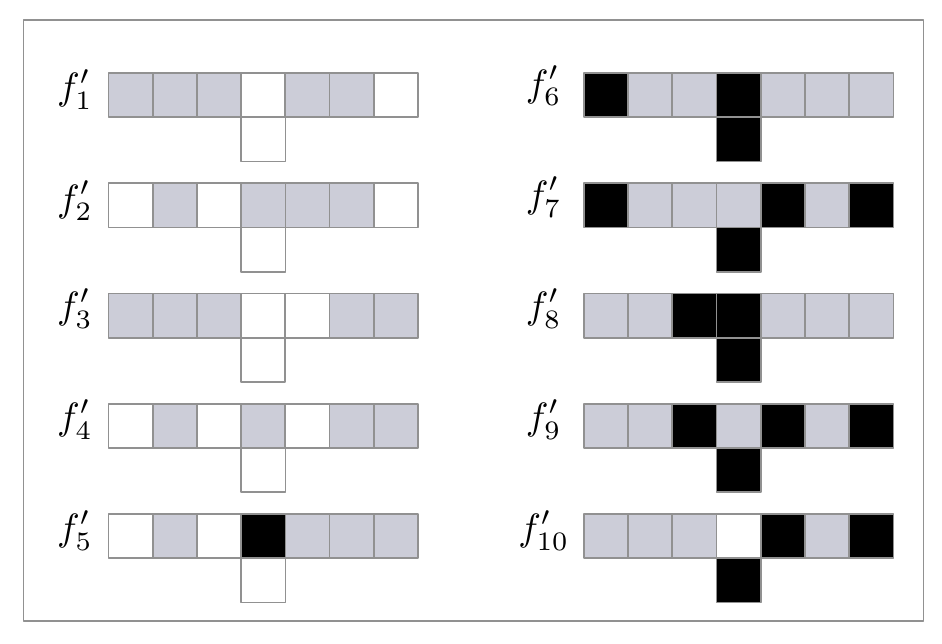}
   \caption{Wildcard schemata of $F_{\textrm{GKL}'}$. Schemata $f'_5$
     and $f'_{10}$ describe all the possible state-changes for a cell;
     all other situations are preservations. Notice that schemata
     $f'_2, f'_4, f'_7$ or $f'_9$, when applicable to a state-change,
     describe LUT entries already described by $f'_5$ or $f'_10$,
     respectively; When prescribing preservations the same schemata
     are captured with less enputs by $f'_1, f_3, f'_6$ or
     $f'_8$. This means that $f'_2, f'_4, f'_7$ or $f'_9$ are not
     essential schemata.}
\label{gkl-e1}
\end{center}
\vspace*{-.2in}
\end{figure}

The wildcard schemata for $F_{\textrm{GKL}'}$ are shown in Figure
\ref{gkl-e1}. 
Note that most schemata are state-preserving: all state-changes in
this automaton are described by two wildcard schemata ($f'_5$ and
$f'_{10}$).
Generations (annihilations) occur only when a cell in state 0 (1)
observes right (left) and right-most (left-most) neighbours in the
opposite state---this is characterized by schemata $f'_5$ and
$f'_{10}$.
In all other situations the a cell $x_i(t+1) = x_i(t)$.

\begin{figure}[!ht]
\begin{center}
   \includegraphics[width=2.9in]{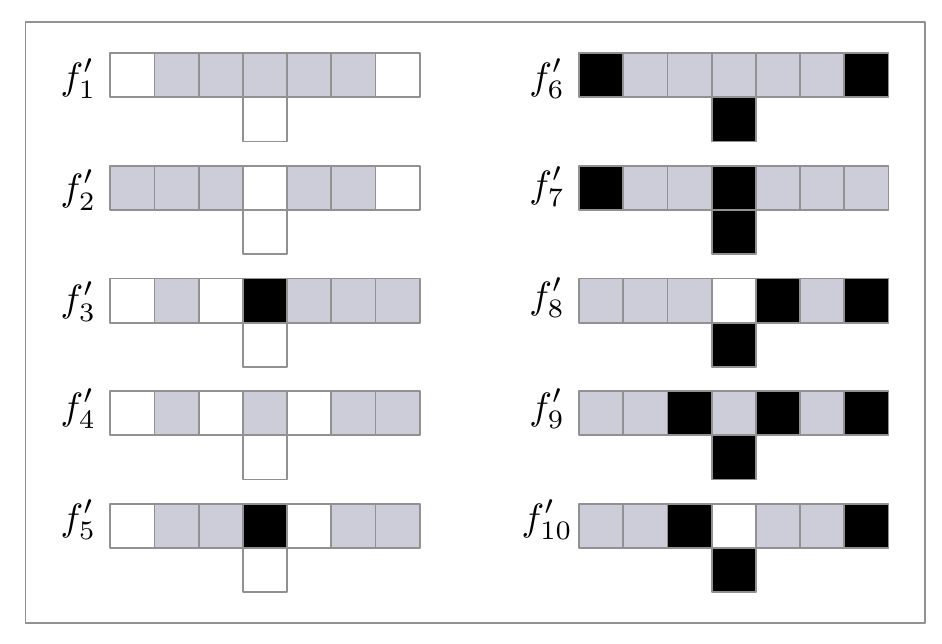}
   \caption{Wildcard schemata for $F_{\textrm{GP}}$. $f'_1$, $f'_3$,
     $f'_5$, $f'_6$, $f'_8$, and $f'_{10}$ describe all the possible
     state-changes for a cell $x_i$; all other situations are
     preservations. Notice that schemata $f'_4$ and $f'_9$, when
     applicable to a state-change, describe LUT entries already
     described by $f'_3$ or $f'_5$ and $f'_8$ or $f'_{10}$,
     respectively}
\label{gp-e1}
\end{center}
\vspace*{-.2in}
\end{figure}

\begin{figure}[!ht]
\begin{center}
   \includegraphics[width=3in]{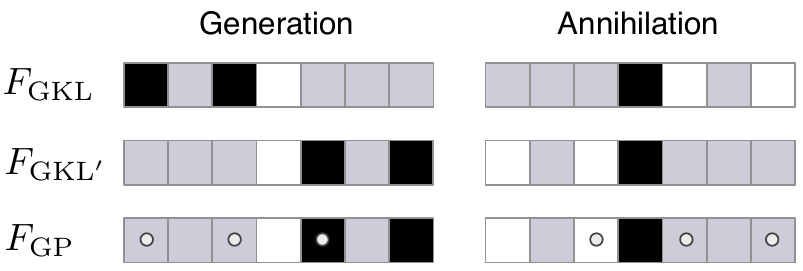}
   \caption{Two-symbol schemata for $F_{\textrm{GKL}'}$
     ($F_{\textrm{GKL}}$) and $F_{\textrm{GP}}$, characterizing every
     possible situation where there are state changes for these
     rules. Note that here we include the state of the updating cell,
     (even if not an enput in the original schemata) to describe the
     known state-changing transitions.} \label{koza-gkl} \end{center} \vspace*{-.2in}
\end{figure}

The wildcard schemata for $F_{\textrm{GP}}$ are shown in Figure
\ref{gp-e1}. In this case, all state changes are
characterized by three wildcard schemata: $f'_1$, $f'_3$, and
$f'_5$ for annihilation, and $f'_6$, $f'_8$, and $f'_{10}$ for generation.
In all other situations the a cell $x_i(t+1) = x_i(t)$.
Using the the two-symbol schemata introduced here, we can appreciate
better the similarity between the two rules.
Figure \ref{koza-gkl} depicts the essential state-changing transitions
in both rules, which we obtained by via two-symbol schema-redescription,
thus removing all redundancy from local dynamics.
We can now, for the first time, understand that $F_{\textrm{GP}}$ is
doing something very similar to, and which includes the behaviour of,
$F_{\textrm{GKL}'}$ at local level.

Both CA rules can be understood by a single schema for generation and their
process-symmetric schema for annihilation. 
Each such schemata is characterized by two enputs. Moreover, one of
the enputs is exactly the same for both CAs: for the generation schema,
both CAs depend on the right-most neighbour being 1, and for
annihilation schemata, both CAs depend on the left-most neighbour
being 0.
The second (group-invariant) enput in $F_{\textrm{GP}}$ is also very
similar to, and indeed contains the second enput for, $F_{\textrm{GKL}'}$.
For the generation schema, $F_{\textrm{GKL}'}$ depends on the
immediate-right neighbour being 1, and for the annihilation schema,
$F_{\textrm{GKL}'}$ depends on the immediate-left neighbour being 0.
When it comes to $F_{\textrm{GP}}$, the second enput is a
group-invariant enput: one of three possible input cells in the
neighbourhood may be specified (set to 0 or 1). 
Specifically, For the generation schema, $F_{\textrm{GP}}$ depends not
only on the right-most neighbour being 1, like $F_{\textrm{GKL}'}$, but
\emph{alternatively}, on the immediate-left, immediate-right \emph{or}
left-most neighbours being 1.
Likewise, for the annihilation schema, $F_{\textrm{GP}}$ depends not
only on the left-most neighbour being 0, like $F_{\textrm{GKL}'}$,
\emph{alternatively}, on the immediate-left, immediate-right \emph{or}
right-most neighbours being 0.

%% This figure (the graphic) shows the prime sign for the GP rule, which is incorrect

\section{Connecting the Levels}
\label{sec:faraway}

The similarity between the local-level descriptions of
$F_{\textrm{GKL}'}$ and $F_{\textrm{GP}}$ we discovered, as well
as their similar performance leads to a natural question:
do the additional regular domains produced by the $F_{\textrm{GP}}$,
play a significant role in its collective computation of the DCT?
The following three facts are known: (1) $F_{\textrm{GP}}$ produces
the same three regular domains as $F_{\textrm{GKL}'}$, plus many
others; (2) the performance difference between the two rules, while
significant, is less than $1\%$; and (3) the number of possible
particles grows very rapidly with number $d$  of domains ($d!/(d-2)!$).
Therefore, it is reasonable to conclude that many---even perhaps most
of the regular domains---produced by $F_{\textrm{GP}}$ must be
redundant; i.e. the particles they produce lead to interactions that
have no effect on information transmission across the lattice.

%%% LMTR: in the below I made substancial changes because what you were
% saying was not quite correct according to the CM framework. A particle
% is defined by a boundary between different domains. So, in the collision
% of the green and orange particles in the figure,  there is a real collision
% that leads to two new particles that have the same exact velocity.
% I don't know what you meant by structure either. Not a good term
% because it can mean so many things. It is sufficient to define particles
% by their velocity (speed and direction).

\begin{figure}[!ht]
\begin{center}
   \includegraphics[width=2.2in]{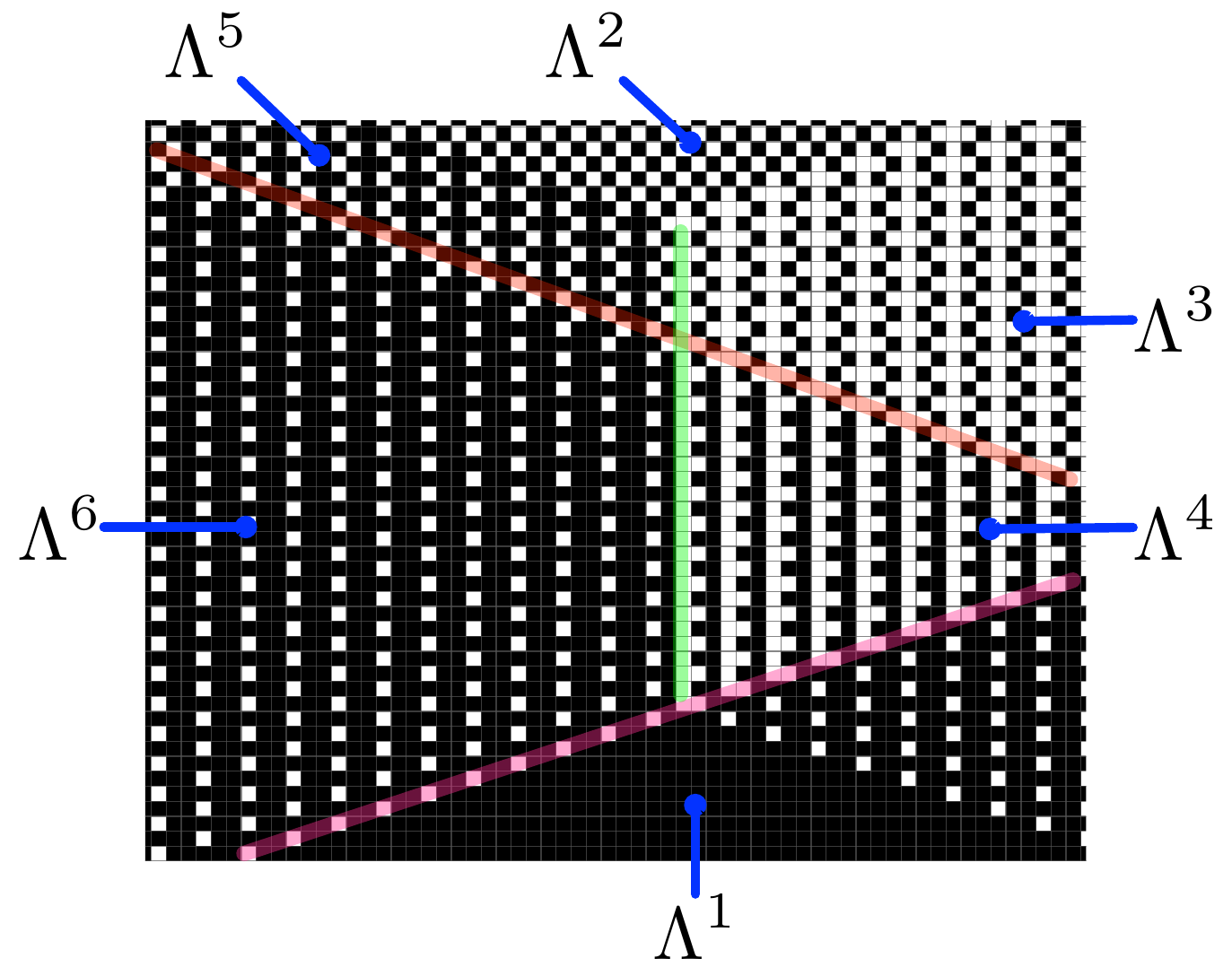}
\vspace*{-.2in}
\caption{ In $F_{\textrm{GP}}$ there are particles that produce equal
  particles after collisions (Highlighted in green (vertical) and
  orange/purple (inclined)} \label{koza_ins}
 \end{center} \vspace*{-.2in} \end{figure}

Support for this conclusion comes from different sources.  One is
simple inspection of the space-time dynamics produced by $F_{\textrm{GP}}$.
Figure \ref{koza_ins} depicts a portion of the space-time dynamics for
$F_{\textrm{GP}}$ shown in Figure \ref{koza-dyn}.
Notice, for example, the collision between the particle formed by
domains $\Lambda^5$ and $\Lambda^3$ (highlighted in green), and the
particle formed by domains $\Lambda^5$ and $\Lambda^6$ (orange). 
This collision, in the CM framework perspective, leads to two new
particles: one formed by domains $\Lambda^6$ and $\Lambda^4$, and
another formed by domains $\Lambda^3$ and $\Lambda^4$.
However, this collision seems to be redundant, leading to particles
that behave exactly as the colliding particles (same velocity,
possibly same structure).
This is highlighted in Figure \ref{koza_ins} by using the same colours
for the particles observed before and after the collision.
%
%Similar situations take place with the other two highlighted
%particles.
%
%Even the encounter between the particle travelling to the right (red)
%and the vertical particle leaves both particles unaffected.
%
% LMR: That is not true; the red particle destroys the green one, and creates
% a new one that you are not coloring (travelling to the left). So what happens here
% is that the collision leads to one particle that behaves the same, and one, that
% behaves differently.
%
This suggests that the (large) set of possible regular domains in the
dynamics of $F_{\textrm{GP}}$ contains sets of domains that are
equivalent, such that collisions among the particles they produce are
redundant.
In the example above $\Lambda^5$ and $\Lambda^6$ seem to be
equivalent, just like $\Lambda^3$ and $\Lambda^4.$

Our observation is supported further when looking at the schemata for
$F_{\textrm{GP}}$.
%
%The appearance of the position-free symbol in generation and assimilation schemata
%for $F_{\textrm{GP}}$, in comparison to the schemata obtained for
%$F_{\textrm{GKL}'}$, means that several neighbourhood configurations (LUT entries) that are preservations in $F_{\textrm{GKL}'}$ become state transitions in
%$F_{\textrm{GP}}$.
%%
%This results in preservation schemata such as $f'_1, f'_2, f'_6$ and
%$f'_7$ (see Figure \ref{gp-e1}), where enputs that are spatially
%distant are sufficient to determine state transitions, disregarding
%the states of even the updating cell in some cases.
%
%%% LMR: I don't think the above is strictly correct... f'_1 and f'_6 are not just
% preservations.... Besides, the gaps that you describe ahead can exist without
% the position-free symbol...
%
We can use the knowledge encoded in schemata, to
better understand the collective behaviour of the CA.
Consider $f'_1$ in $F_{\textrm{GP}}$, which enforces that a
cell $x_i$ in any state whose left- and right-most neighbours are 0
will transition to, or stay 0 at $t+1$ (see Figure \ref{koza-gkl}).
This leaves a wide ``gap'' of five cells in-between the 0-cells, which
can be in any state, without affecting the state-transition of $x_i$
to 0.
When this pattern is repeated in the lattice, i.e. \emph{every
  $6^{\textrm{th}}$ cell is 0}, using the two-symbol schemata obtained
instead of the original LUT, we can compute the resulting collective
dynamics with \emph{incomplete} information.
Indeed, one key advantage of schema redescription is that it allows
the simulation of the dynamics of automata networks from ICs where
only the state of some cells is specified, while the state of the
others is \emph{unknown}.

\begin{figure}[!ht]
\begin{center}
   \includegraphics[width=3.5in]{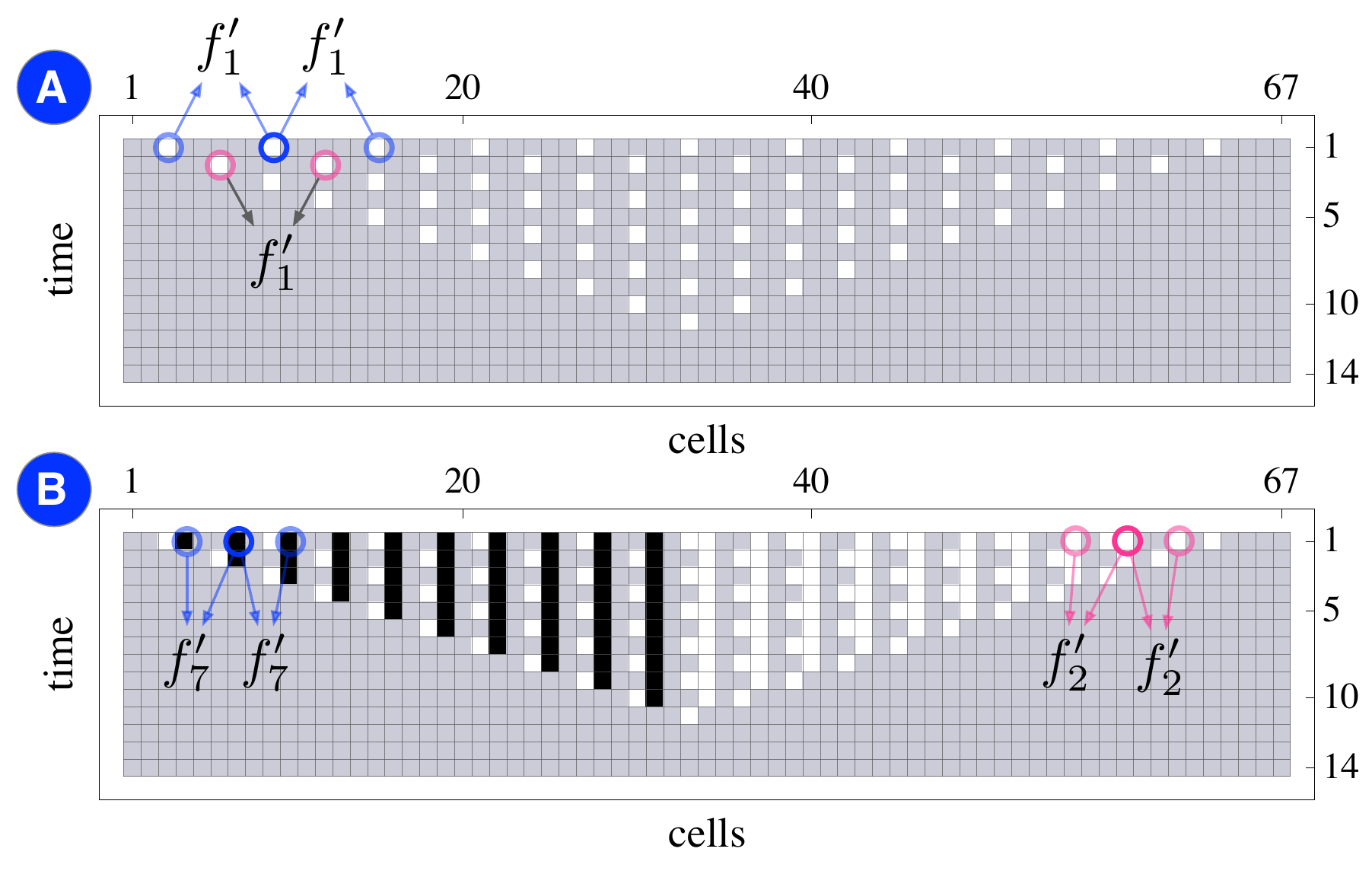}
\vspace*{-.2in}
\caption{Dynamically decoupled dynamics of patterns in
  $F_{\textrm{GP}}$. See text for description. }
\label{koza_dec}
\end{center}
\vspace*{-.15in}
\end{figure}

Figure \ref{koza_dec} (A) depicts the \emph{known} dynamics given
partial lattice information, where every sixth cell is 0 (except at
the lattice borders) and the state of grey cells is \emph{unknown}.
This allows us to know how the initial input pattern will propagate
in the CA dynamics.
In this case, our lack of knowledge propagates in the lattice, and our
partial knowledge of the dynamics eventually disappears completely.
In \ref{koza_dec} (B) we show that different such patterns can
co-exist independently, in the previous unknown segments of the
lattice. The difference from (A) is that every third cell in the left
half of the initial lattice is set to 1, and the third and
every third cell after that is set to 0, in the right half of the
initial lattice (these conditions match schemata $f'_7$ and $f'_2$ respectively).

One very interesting observation from computing dynamics with
incomplete information, is that many patterns (such as those in Figure
\ref{koza_dec}) are \emph{dynamically decoupled} in 
$F_{\textrm{GP}}$.
This means that the space-time patterns produced from schemata $f'_1,
f'_2$ and $f'_7$ in $F_{\textrm{GP}}$ are updated independently: they
function as building blocks that can switch places, producing
different domains (from the CM perspective) that in reality are formed
of some of the same sub-patterns (defined by the same schemata).
This is clear in Figure \ref{koza_dec} (B) where the patterns from
schemata $f'_1, f'_2$ and $f'_7$, coexist independently of one
another.
Since some of the schemata for $F_{\textrm{GP}}$ have large ``gaps''
wherein the states of the cells are irrelevant, if particles collide
inside those gaps, some of the decoupled underlying patterns are
unaffected, leading to equivalent domains and redundant collisions
such as what we apparently observe in Figure \ref{koza_ins}.

In contrast, rule $F_{\textrm{GKL}'}$ does not easily lead to dynamic
decoupled patterns and redundant collisions, since its schemata do not
leave large enough ``gaps''  in the lattice for other patterns to
occur.
Therefore, $F_{\textrm{GKL}'}$ presents much fewer domains than rule
$F_{\textrm{GP}}$. But since the domains obtained by rule
$F_{\textrm{GP}}$ are the result of schemata with a large amount of
canalization leading to large ``gaps'' that can be in any state, many
of the obtained distinct domains are  dynamically
equivalent.

\begin{figure}[!ht]
\begin{center}
   \includegraphics[width=2.9in]{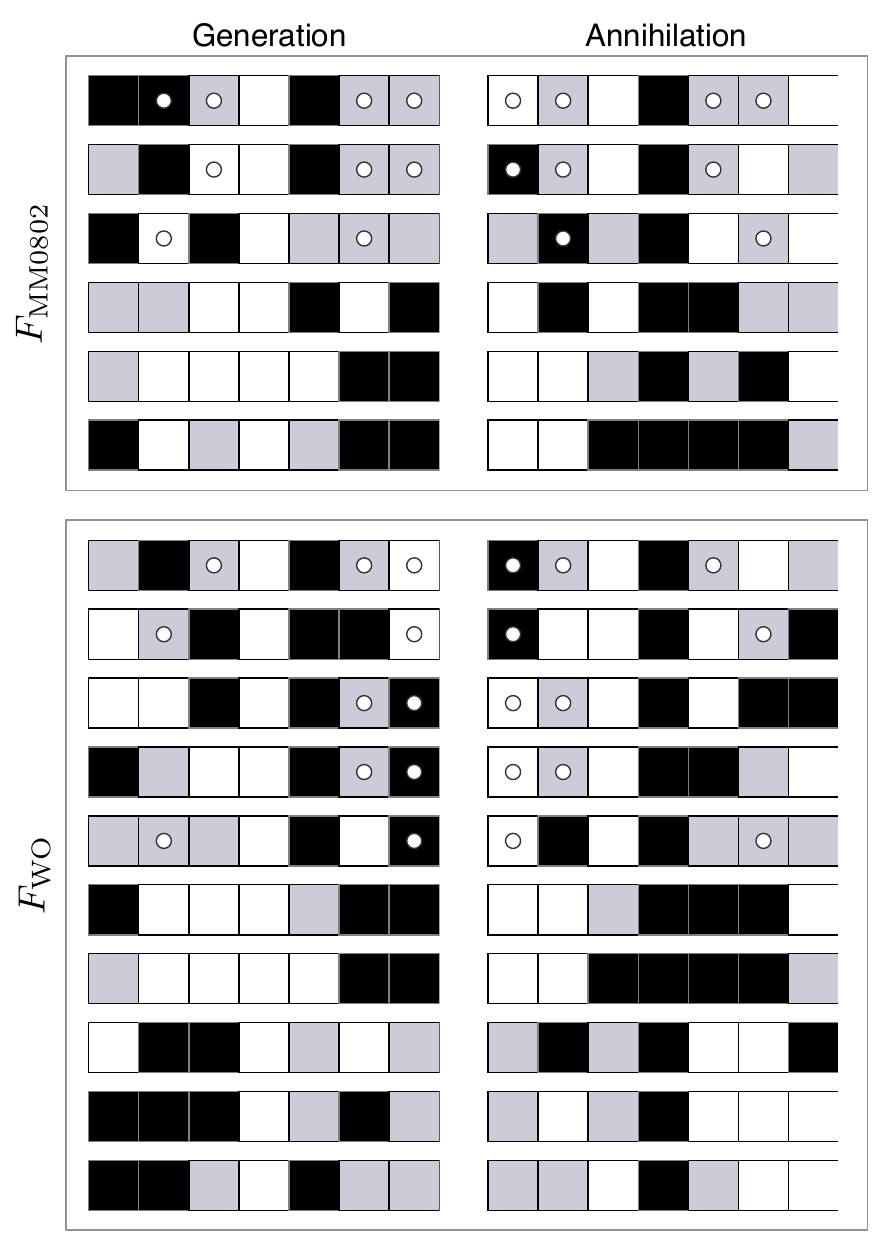}
   \caption{2-symbol schemata of $F_{\textrm{MM0802}}$ and $F_{\textrm{WO}}$, characterizing every possible situation where there are state changes for this rule.}
\label{mm0802_wo}
\end{center}
\vspace*{-.2in}
\end{figure}

This observation is also supported by the schema redescription of
additional, better-performing rules for the DCT: $F_{\textrm{MM0802}}$
 \cite{Marques-Pita:2008cr} and $F_{\textrm{WO}}$
 \cite{Wolz:2008fk}, with performances $\mathcal{P}^{10^5}_{149}(F_{\textrm{MM0802}})
 \approx 0.845$ and $\mathcal{P}^{10^5}_{149}(F_{\textrm{WO}}) \approx
 0.889$. 
 The two-symbol (state-changing) schemata for these rules are depicted
 in Figure \ref{mm0802_wo}.
As we can see, for the DCT, increasingly better-performing rules
become ``less canalized''.
This means that they are very sensitive to noise in ICs (as we would
expect for the DCT): bit-flips or incomplete state-specification have a
high chance of leading to wrong classification (in DCT terms).
Since noise is not used in their performance computation and fitness
evaluations, these rules find advantages in becoming increasingly
specific to the initial conditions they respond differently to.
Therefore, they have a higher number of schemata with more enputs and
therefor less canalization to describe their behaviour.
The dynamics of these CAs do not produce the large
numbers of domains as observed in $F_{\textrm{GP}}$. \footnote{
  redescriptions
  and dynamics of all the CAs in this paper are available from
  \texttt{http://cnets.indiana.edu/groups/casci/mrecinca/ca}}

\section{Discussion}
\label{sec:discussion}

Two-symbol schema redescription allowed us for the first time to
understand that despite dramatically distinct collective dynamics
(emergent behaviour), the $F_{\textrm{GP}}$ rule is a more general
version of the $F_{\textrm{GKL}'}$ rule.
The statistically significant, albeit small, performance improvement
of $F_{\textrm{GP}}$ over  $F_{\textrm{GKL}'}$, is gained by
converting a single enput, into a group-invariant enput.
In other words, $F_{\textrm{GP}}$ introduced ambiguity to the
dynamical behaviour \cite{klir99gis} of $F_{\textrm{GKL}'}$, by
allowing the state-transition to optionally occur for a larger set of
configurations of a cell's neighbourhood.
This redundancy in the form of ambiguity is quite effectively captured
by our two-symbol schema redescription with the position-free symbol.

It should be emphasized that until this point, and especially when
looking only at the emergent behaviour of these rules, their deep
similarity was unknown. With schema redescription of local-level
dynamics, not only can we uncover the similarity, but we can
explain it.
It also makes sense that $F_{\textrm{GP}}$ was obtained via genetic
programming \cite{Andre96b}.
This rule was probably evolved via minor changes to a tree encoding of
$F_{\textrm{GKL}}$; it would need simply to substitute a tree terminal
for a sub-tree with a disjunction node for 3 terminal edges, plus a
mirror operation to go from $F_{\textrm{GKL}}$ to $F_{\textrm{GKL}'}$.

We can also see that both rules are extremely \emph{canalized}: every
possible state-change is controlled by two enputs, disregarding the
states of all other neighbouring cells. 
But the $F_{\textrm{GP}}$ is canalizing in yet another way: it
depends, alternatively, on the state of one of the input cells in a
group-invariant enput.
These group-invariant (collective) enputs can be formally described as
a disjunction of alternative input variables (logical literals) and captured in
schemata redescription by the additional position-free symbol.
This additional form of canalization in Boolean functions, and its
effect on determining collective dynamics in automata networks used to
model genetic regulation and signalling networks is a topic we will
explore further, elsewhere.

Another key idea we will explore in future work is how to use the
explanatory power of schema redescription at the local level of
Boolean LUTs, to link local to global behaviour in automata networks.
From what we presented here, we can already see that the
$F_{\textrm{GP}}$ rule must necessarily lead to some equivalent
domains and redundant particle interactions.
Since it contains group-invariant enputs, it canalizes distinct and
alternative neighbourhood input patterns to the same transition.
Therefore, domains that seem to be distinct at the
collective space-time behaviour level, can in reality be equivalent
and not involved in transmitting novel information through the
lattice, merely maintaining the same information in different-looking
domains.
In a forthcoming paper we will build up from schema redescription at
the local level, and the computation of collective dynamics from
incomplete information, to automatically identify domains (relevant
and redundant), particles, and particle interactions at the collective
dynamics level.

As a final topic of discussion, the present work moves us to consider
that an exclusively emergentist approach to complex systems dynamics
is incomplete.
The induction of recurring patterns in dynamics is not a substitute
for uncovering the lower-level mechanisms that actually dictate the
dynamics. 
A full account of complex systems requires attention to
local and global dynamics.
In a sense, too much attention to the ``spots'' or ``stripes'' we
observe at the emergent level of behaviour, can mask that underneath
it all there may be a very similar animal.

\section*{Acknowledgments}

This work was supported by Funda\c{c}\~ao para a Ci\^encia e a
Tecnologia (Portugal) grant 36312/2007.  We thank the FLAD
Computational Biology Collaboratorium at the Gulbenkian Institute
(Portugal) for hosting and providing facilities used for this
research. We also thank Indiana University for
providing access to its computing facilities.


\cleardoublepage
\bibliographystyle{IEEEtran}
\bibliography{./MMPRefs}

\end{document}